\documentstyle{aipproc}  
\begin{document}
\begin{flushright}
UH-511-988-01 \\
September 2001
\end{flushright}
\title{Chiral Symmetry and Scalars\footnote{Parallel Session Talk at Hadron2001-Protvino}}
\author{S.F.~Tuan}
\address{Department of Physics, University of Hawaii at Manoa \\ 
Honolulu, HI 96822-2219, U.S.A.}
\maketitle
\begin{abstract}
The suggestion by Jaffe that if $\sigma$ is a light $q^{2}\bar{q}^{2}$ state
$0^{++}$ then even the fundamental chiral transformation properties of the
$\sigma$ becomes {\bf unclear}, has stimulated much interest. Adler pointed
out that in fact the seminal work on chiral symmetry via PCAC consistency,
is really quite consistent with the $\sigma$ being predominantly
$q^{2}\bar{q}^{2}$. This interpretation was actually backed by subsequent
work on effective Lagrangian methods for linear and non linear realizations.
More recent work of Achasov suggests that intermediate four-quark states
determine amplitudes involving other scalars $a_{0}(980)$ and $f_{0}(980)$
below 1 GeV, and the report by Ning Wu that study on $\sigma$ meson in
$J/\psi \rightarrow \omega\pi^{+}\pi^{-}$ continue to support a non $q\bar{q}$
$\sigma$ with mass as low as 390 MeV. It is also noted that more recent
re-analysis of $\pi K$ scattering by S. Ishida {\em et al.} together with
the work of the E791 Collaboration, support the existence of the scalar
$\kappa$ particle with comparatively light mass as well.
\end{abstract}

In an intriguing paper Jaffe\cite{REF1} pointed out that the QCD ``Breit
Interaction'' summarized by an effective Hamiltonian acting on the quarks'
spin and color indices,
\begin{eqnarray*}
H_{eff} \ \propto - \sum_{i \neq j } \ 
\stackrel{\lambda}{\sim} _i .
\stackrel{\lambda}{\sim} _j\vec{\sigma}_i . \vec{\sigma}_j
\end{eqnarray*}
affirm earlier work\cite{REF2} that $f_{0}(980), a_{0}(980), \sigma(560)$,
and $\kappa(900)$ scalars make a nonet with mass spectrum, decay couplings
and widths that look qualitatively like $\bar{q}\bar{q}qq$ system. Alford
and Jaffe\cite{REF3} raised the pertinent question that if light 
$\bar{q}^{2}q^{2}$ states are, in fact, a universal phenomenon below 1 GeV,
and if $\sigma$ is predominantly a $\bar{q}^{2}q^{2}$ object, then the 
chiral transformation properties of the $\sigma$ have to be re-examined.
The $\pi$ and the $\sigma$ are usually viewed as members of a (broken)
chiral multiplet. In the naive $\bar{q}q$ model both $\pi$ and $\sigma$
are in $(1/2,1/2) \oplus (1/2,1/2)$ representation of $SU(2)_{L} \otimes
SU(2)_{R}$ before symmetry breaking. In a $\bar{q}^{2}q^{2}$ model, as in
the real world, the chiral transformation properties of the $\sigma$ are
{\bf not clear}.

There remains a body of recent literature\cite{REF4} which retains in 
essence the $\bar{q}q$ model for the $\sigma$ meson. For instance 
T\"{o}rnqvist {\em et al.}\cite{REF4} used chiral-symmetry constraints in
their study. Chiral symmetry constraints have been discussed by for 
instance Oller\cite{REF5} where it is said that the range of applicability
of chiral constraints could be enlarged up to around 0.8 GeV. Because of
the model dependence of experimental analysis, current wisdom suggests that
$\sigma$ has mass between 400 to 700 MeV and hence the use of chiral
constraints would appear to be valid. However in the context of 
T\"{o}rnqvist {\em et al.}\cite{REF4} the 4q $\bar{q}^{2}q^{2}$ scheme is
not easy to combine with chiral symmetry constraints, which are 
crucial to their work. Indeed for weak interactions, their (chiral) results
are the same as the strong interaction quark-level linear $\sigma$ model
(LSM) $\bar{q}q$ scheme in one-loop order together with the electromagnetic
(LSM) analogue\cite{REF6}. T\"{o}rnqvist\cite{REF6} expressed further
concern that 4q or rather 2 meson models and chiral symmetry, the chiral
symmetry can of course be imposed in a model like the linear $\sigma$ model
(LSM), but then all states $\sigma$, $f_{0}$, $a_{0}$ and the {\bf pion}
would be basically 4q states! We shall return to this concern later on in
this paper.

On an optimistic note, Adler\cite{REF7} pointed out that in the original
PCAC Consistency Condition paper\cite{REF8}, when analysed for the pion-
pion scattering case, led to the conclusion that there had to be a broad
low energy pion-pion scattering resonance. This is then quite consistent
with the $\sigma$ being predominantly $q^{2}\bar{q}^{2}$. Secondly, the 
numerical estimates of the ``sigma term'' from current algebra\cite{REF9},
assuming it is $q\bar{q}$ [or $(3,\bar{3}) + (\bar{3},3)]$ were always an
embarrassment, since they were generally off by a factor of two whereas
other things worked much better than that (typically of order 10\% or 
less\cite{REF8}). This again is quite consistent with the dominant spectral
weight not being in the $q\bar{q}$ channel. Third, in Zumino's 1970 Brandeis
lectures\cite{REF10} on effective Lagrangian methods, he discusses nonlinear
realizations on pp. 451-454 (see also pp. 481-483, 485); he first describes
the linear realization of the $\sigma$ model, stating that $\sigma$.....
(is the field) of a scalar isoscalar $\pi-\pi$ resonance. He then shows
how by a redefinition the same low energy results arise from a nonlinear
transformation involving the redefined pion field only; in this nonlinear
transformation, $\vec \pi^{2}$ plays a role analogous to that played by
$\sigma$ in the linear case. So again, it is expected that the $\sigma$
should be a two pion state, and hence not surprisingly that it is dominantly
$q^{2}\bar{q}^{2}$.

Jaffe\cite{REF11} expanded on his understanding (or lack thereof) of the
role of $\sigma$ in chiral symmetry\cite{REF3}. Since chiral SU(2) symmetry
is spontaneously broken, the physical particles do not have to transform
as irreducible representations of $SU(2) \times SU(2)$. There is a prejudice
(originating in the quark model?) that the pion transforms like $\bar{q}q$,
and an even less well justified prejudice that the $\sigma$ transforms in
the same way as the pion. However, there does not exist any good reason to
think that the transformation properties of the $\sigma$ are linked to
those of the pion when $SU(2) \times SU(2)$ is spontaneously broken. Perhaps
another way of saying the same thing\cite{REF12} is that chiral symmetry
does not mesh well with either constituent quarks nor with QCD's current
quarks, hence chiral symmetry does not require multi-quark states to fuse
into a $q\bar{q}$ state as originally thought. 

Experimental evidence for the existence of the scalar $\sigma$ at the low 
mass value of 390 MeV with total width of order 282 MeV has been recently
reported by Ning Wu\cite{REF13} based on the study of $\sigma$ particle in
$J/\psi \rightarrow \omega\pi^{+}\pi^{-}$ from $7.8 \times 10^{6}$ BESI
$J/\psi$ data. There is also the newly reported\cite{REF14} $\sigma(\pi\pi)$
scalar resonance with a $\sigma$ mass and width of $478 \pm 24 \pm 17 
MeV/c^{2}$ and $342 \pm 42 \pm 21 MeV/c^{2}$. Indeed recent re-analysis of the
$\pi\pi$ scattering data by S. Ishida {\em et al.}\cite{REF15} shows evidence
for the existence of $\sigma$ with comparatively light mass also. This same
scattering data\cite{REF15} for $\pi K$ also showed evidence for the 
existence of the $\kappa$ particle also of relatively light mass. This is
corroborated again by the newly reported\cite{REF14} $\kappa(K\pi)$ scalar
resonance with a $\kappa$ mass and width of $815 \pm 30 \ MeV/c^{2}$ and
$560 \pm 116 \ MeV/c^{2}$. However Achasov\cite{REF16} has cautioned that
information on these scalars can be obtained only in strongly model
dependent ways up to now. It seems reasonable that together with the status
of $f_{0}(980)$ and $a_{0}(980)$ rather carefully analysed by Achasov and
Gubin\cite{REF17} {\bf we do nevertheless have a nonet of $q^{2}\bar{q}^{2}$
scalars below 1 GeV}, though the mass and width of some of these scalars
remain to be pinned down more precisely. Coming back to a more theoretical
understanding of the situation, Achasov\cite{REF16} reassured that 
T\"{o}rnqvist's fear\cite{REF6} that the pion also may end up as a 4q state
is strongly overstated. The point is that one can not say that a field 
contains a fixed number of quarks. It is approximately true only in some
energy (virtuality) region. For example, when virtualities of $\sigma$ 
states have the order of the pion mass they show themselves as two-quark
states, the chiral partners of pions, but when virtualities of $\sigma$
states are of the order of 1 GeV (remember a $\sigma$ of mass 700 MeV 
remains in the acceptable range), they can show themselves as four-quark
states. Jaffe\cite{REF11} elaborated further that the ``quark content'' of
a particular meson is a heuristic concept at best. In some contexts the
pion appears to be a $q\bar{q}$ state (for example as a member of an SU(3)
meson octet); in others it appears to be a ``wave on the chiral vacuum'',
which would be a coherent state in the Bogoliubov sense, including 
arbitrarily high numbers of $q\bar{q}$. The point is that the $\sigma$,
the $f_{0}(980)$, and $a_{0}(980)$ has always been that the principal
features of their mass spectrum, couplings to pseudoscalars, and to 
electromagnetic fields, are well described by a dominant $qq\bar{q}\bar{q}$
content. Hence there is agreement with Achasov that the quark content can
be regarded as ``virtuality'' dependent. Jaffe\cite{REF11} also pointed
out that his understanding of Jona Lasinio/Nambu spontaneous symmetry
breaking where
\begin{equation}
\sigma = \sqrt{1-\vec{\pi}^2 / f^2_\pi}
\end{equation}
\begin{equation}
= 1- \frac{\vec{\pi}^2}{2 f^2_\pi}   \  +  \cdot \cdot \cdot \cdot \cdot
\end{equation}
is in fact the same as that of Zumino\cite{REF10} who discussed spontaneous
symmetry breaking in the non linear realization case as
\begin{equation}
\delta \vec{\pi} = 2 \vec{\alpha} \sqrt{\kappa^2 - \vec{\pi}^2}
\end{equation}
where $\kappa$ = $(1/2)f_{\pi}$. Expansion of the r.h.s. of (3) in terms of 
[$\vec \pi^{2}$/$\kappa^{2}$], one would get something very similar to the
r.h.s. of (2) up to a multiplicative factor. Hence Jaffe is in agreement
with Adler\cite{REF7}.

We have certainly come a long way from the traditional naive quark model
classification of hadron states of some 35 years ago\cite{REF18}. For some
trained in the traditional approach like myself, what is described above
comes as a surprise bordering on shock. Hence the opportunity to air out
these concerns at Hadron2001 is much appreciated.  (During the
discussions after this talk, Professor J. Schechter pointed out the work
of the Syracuse group\cite{black} which addressed quantitatively some of
the issues presented here.)

I wish to thank my scientific colleagues Kolia Achasov, Steve Adler, 
Bob Jaffe, and
Nils T\"{o}rnqvist for very helpful communications and discussions. This
work was supported in part by the U.S. Department of Energy under Grant
DE-FG-03-94ER40833 at the University of Hawaii at Manoa.

\end{document}